\documentclass{ws-procs9x6}

\begin{document}

\title{Efficient Evaluation of Effective Action in Radial Backgrounds}

\author{Hyunsoo Min\footnote{E-mail: hsmin@dirac.uos.ac.kr. This work is supported in part by University of Seoul 2007 Research Funding Program.}}

\address{Department of Physics, University of Seoul,\\
Seoul, 130-743, Korea}

\begin{abstract}
Recently  a new
caculational scheme for effective actions in radial background fields was developed. The
effective action is expressed as an infinite sum of partial-wave
contributions, using the rotational symmetry of the system.
The sum becomes convergent after proper regularization and
renormalization, but the rate of convergence is rather slow. 
We introduce a systematic way of accelerating the rate of convergence. This method is based
on a radial WKB series in the angular momentum cut-off. We demonstrate 
the power of this scheme
by applying it to the calculation of instanton determinant in QCD.
\end{abstract}


\bodymatter

\section{Introduction}\label{abs:sec1}
The one-loop effective action plays a central role in quantum field theories. 
However, the explicit evaluation of it is usually very difficult  and 
analytic results are known only in  limited cases. 

Recently there has been a significant progress in this problem when background fields have radial symmetry,  using the partial wave analysis\cite{idet,radial1}.
Introduction of a cut-off in the partial wave sum  and separation of the sum into two parts is a key of the idea. 
A radial WKB expansion which is uniformly valid for the large angular momentum part is developed.  Proper renormalization counterterms are taken into account there. 
Combination of the leading terms of this WKB expansion and the contributions from the low angular momentum part provides us a finite renormalized value in the limit of large cutoff value. 
It also turns out that the inclusion of higher order terms in the uniform radial WKB expansion greatly improves the rate of convergence in the infinite sum\cite{radial2}.  

In this work, we present a general scheme of this method 
and some results for the one-loop effective action in the case of QCD instanton.

\section{Partial Wave Method}
The one-loop effective action for a complex scalar field is defined in terms of functional determinants formally as
\begin{eqnarray}
\Gamma = \ln\left( \frac{\det[{\cal M}+m^2]}{\det[{\cal M}^{\rm free}+m^2]} \right)  
\label{effaction}
\end{eqnarray}
where ${\cal M}=-\partial^2 +V(r)$, ${\cal M}^{\rm free}=-\partial^2$.
Here $V(r)$ is a radial potential vanishing sufficiently fast at infinity. Using the partial wave analysis, the effective action can be written as
\begin{eqnarray}
\Gamma= \sum_{l=0}^\infty g_l\Omega_l, \quad \Omega_l=\ln\left( \frac{\det[{\cal M}_l+m^2]}{\det[{\cal M}_l^{\rm free}+m^2]}  \right).
\label{detlsum}\end{eqnarray}
Here $l$ denotes the angular momentum quantum number appropriate to each partial wave and
$ g_l={(2l+d-2)(l+d-3)!}/{l!(d-2)!}$
is the degeneracy factor\cite{dunnekirsten}. The associated radial differential operator${\cal M}_l$ is given by ${\cal M}_l=-\partial^2+{\cal V}_l$ with the effective potential
\begin{eqnarray}
{\cal V}_l =  \frac{(l+\frac{d-3}{2})(l+\frac{d-1}{2})}{r^2} +V(r),
\end{eqnarray}
and ${\cal M}_l^{\rm free}={\cal M}_l$ with $V=0$.

The individual radial determinant ratio $\Omega_l$ in (\ref{detlsum}) can be evaluated easily  by using the  Gel'fand-Yaglom method\cite{gy}
\begin{equation}
\Omega_l = \lim_{r\to\infty} \ln \frac{\psi_l(r)}{\psi_l^{\rm free}(r)},
\end{equation}
where the wave functions $\psi_l(r)$ and $\psi_l^{\rm free}(r)$ are the solutions of  $({\cal  M}_l+m^2)\psi_l=0$ and $({\cal  M}_l^{\rm free}+m^2)\psi_l^{\rm free}=0$ respectively and both of them behave as $r^{l+(d-2)/2}$ near $r=0$.

The infinite sum in (\ref{detlsum}) is formally divergent. This problem is related to renormalization. There exists an elegant way to extract the renormalized quantity $\Gamma_{\rm ren}$ from $\Gamma$\cite{idet, dunnekirsten}. Another problem is the slow rate of convergence of the $l$-sum.
Both of these problems can be solved by splitting the sum into two pieces: the low angular momentum part $\Gamma_{\rm L}$ and the high angular momentum part $\Gamma_{\rm H}$ as
\begin{eqnarray}
\Gamma_{\rm ren} = \Gamma_{\rm L} +\Gamma_{\rm H}= \sum_{l=0}^L g_l\Omega_l + 
(\sum_{l > L}^\infty g_l \Omega_l+\delta\Gamma).
\label{lcut}  
\end{eqnarray}
Here we introduce a cut-off $L$ and $\delta\Gamma$ denotes the `conventional' renormalization counterterm. 
Each $\Omega_l$ in the part $\Gamma_{\rm L}$ can be evaluated using the above Gel'fand-Yaglom method. Since $\Omega_l$ behaves like $\sim 1/l$ for large $l$ and $g_l$ increases as $l^{d-2}$, $\Gamma_{\rm L}$ behaves like $L^{d-2}$ for $d\geq2$ in the large $L$ limit. (This reveals the divergent structures in the formal expression in (\ref{detlsum})). 
As for the part $\Gamma_{\rm H}$, we can evaluate it {\em analytically} in a uniform asymptotic series of the form
\begin{equation}
\Gamma_{\rm H}=\int_0^\infty dr \left( Q_{\rm log} +\sum_{n=2-d}^\infty Q_{-n}L^{-n}\right), \label{asympL}
\end{equation}
where $Q_{-n}$'s may have an implicit $L$ dependence of $O(L^0)$ and $Q_{\rm log}$ behaves as $O(\ln L)$ in the large $L$ limit.  One can find explicit forms of the $Q$'s with derivation in a recent work\cite{radial2,hurmin}. 

Now let us consider the case with very large value of $L$.
Note that, as $L\to\infty$, unsuppressed terms in the expansion (\ref{asympL}) may grow but they match precisely the divergences coming from $\Gamma_{\rm L}$ with the opposite sign. Hence, combining these two and taking the $L\to\infty$ limit yields $\Gamma_{\rm ren}$:
\begin{equation}
\Gamma_{\rm ren}= \lim_{L\to\infty}\left[\Gamma_{\rm L} + \Gamma^{(1)}_{\rm H}\right], \quad \Gamma^{(1)}_{\rm H}= \int_0^\infty dr \left( Q_{\rm log}  +\sum_{n=0}^{d-2} Q_{n}L^{n}\right). \label{subtracted}
\end{equation}
In principle, one can use this expression to obtain the renormalized effective action. 
But we still have a practical problem related with the convergence rate.

With a finite value of $L$, we can write the following formula for $\Gamma_{\rm ren}$:
\begin{equation}
\Gamma_{\rm ren}= \Gamma_{\rm L} + \Gamma^{(1)}_{\rm H}+\int_0^\infty dr  \sum_{n=1}^{N} Q_{-n}\frac{1}{L^{n}} +O\left(\frac{1}{L^{N+1}}\right),  \label{truncated}
\end{equation}
where $N$ refers to the order of truncation. In this formula the error is indicated by the last term and it is totally under control. It is apparent that we get a more accurate value of $\Gamma_{\rm ren}$ by taking into account more $\frac{1}{L}$-suppressed terms for a given value of the cutoff $L$. 
In the subsequent sections, by applying this method to the evaluation of the instanton determinant, we demonstrate the power of our method.

\section{Instanton determinant with $m=0$}
We apply our method  to the case of the instanton determinant with $m=0$ where an exact computation is possible\cite{thooft}. Now the partial derivative $\partial_\mu$ in ${\cal M}$ should be replaced with
$D_\mu = \partial_\mu-iA_\mu$. Here the background field $A_\mu=A_\mu^a \tau^a/2$ is an SU(2) single instanton solution  and $ A_{\mu}^a(x) = 2 \eta_{\mu\nu a} x_\nu f(r)$ with $f(r)={1}/{(1+r^2)}$, taking the size parameter $\rho=1$.
It is convenient to introduce the total angular momentum $\vec{J}=\vec{L}+\vec{T}$.
Then each partial wave is labeled by $(l,j)$ with $j=l\pm1/2$ and  $l$ takes half integral values. 
The effective potential has the form:
\begin{equation}
{\cal V}_{l,j}= \frac{(2l+\frac{1}{2})(2l+\frac{3}{2})}{r^2}
 +4f(r) [ j(j+1)-l(l+1)-{3}/{4}]+3r^2f(r)^2.
\end{equation}

We may exactly solve the radial ODE, ${\cal M}_{(l,j)}\psi_{l,j}=0$ and get a master formula of $\Omega_l$:
\begin{equation}
 \Omega_l=\Omega_{l,l+\frac{1}{2}} +\Omega_{l+\frac{1}{2},l} =\ln\frac{2l+1}{2l+2}.
\end{equation}
The WKB large-$L$ expansion of $\Gamma_{\rm H}$ is evaluated as
\begin{eqnarray}
\Gamma_{\rm H}&=&\Gamma_{\rm H}^{(1)} -\frac{1}{6L}+ 
 + \frac{119}{1440 L^2} - \frac{13}{240 L^3}
+\frac{ 1597}{40320 L^4}- \frac{103}{3360 L^5}+O(L^{-6}), \nonumber \\
 \Gamma_{\rm H}^{(1)}&=&\frac{127}{72}  - \frac{1}{3}\ln2 - \frac{1}{6} \ln L+ 4 L + 2 L^2.
\end{eqnarray}
Then the renormalized effective action is (setting the renormalization parameter $\mu=1$)
\begin{eqnarray}
\Gamma_{\rm ren}=\lim_{L\to\infty} ( \sum_{l=0,{1}/{2},\cdots,L}
g_l\ln\frac{2l+1}{2l+2} + \Gamma_{\rm H}^{(1)}  )\equiv \alpha,
\end{eqnarray}
with
$\alpha=-(5/72) -1/6 \ln 2 - 2 \zeta'(-1)=0.145873312863\ldots$.
Taking a finite value of $L$, we get the approximate value of the effective action:
\begin{eqnarray}
\Gamma_{\rm ren}(L)=\Gamma_{\rm L} +\Gamma_{\rm H} ({\rm truncated}).
\end{eqnarray}
When $L=20$, comparing the
exact and approximate values we find that the difference is just $3.7\times 10^{-10}$.
This result clearly shows the elegance of our method.

\section{Numerical values and asymptotic expansions: $m\neq0$}
\begin{figure}
\includegraphics[scale=1.3]{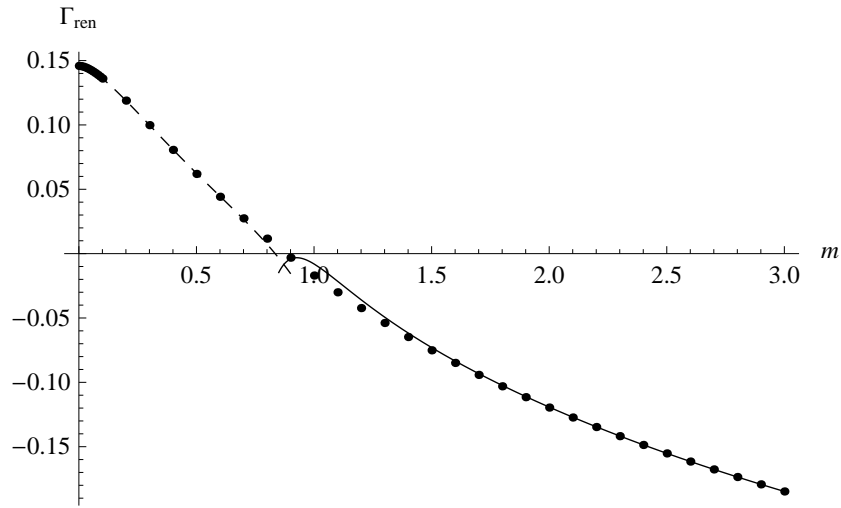}
\caption{Plot of our small mass expansion (dashed line)  and the large mass expansion (solid line) together with the exact numerical result (dots).} \label{fig1}
\end{figure}

When $m\neq0$ it is no longer possible to find a master formula for $\Omega_l$. Related ODE's must be 
solved in a numerical way.
There are other approximate but analytic expressions for the effective action.
The large mass expansion is directly obtained from 
the Schwinger-DeWitt expansion as described in the work\cite{kwon}
\begin{eqnarray}
\Gamma_{\rm ren}=-\frac{\ln m}{6}-\frac{1}{75 m^2}-\frac{17}{735 m^4}+\frac{232}{2835 m^6}-\frac{7916}{148225 m^8}+\cdots
\end{eqnarray}
In a recent paper\cite{hurjin}, the small mass expansion is reported as
\begin{eqnarray}
\Gamma_{\rm ren}&=&\alpha(1/2)+\frac{m^2}{2}(\ln m+\gamma+1/2-\ln2)-\frac{1}{4}m^4\ln^2 m \nonumber\\
&&+m^4( \frac{\ln m}{2}[1/2-\gamma+\ln2] -0.382727) +O(m^6)
\end{eqnarray}
In Fig. \ref{fig1}, we have plotted these two expansions and the exact numerical values of the effective action
$\Gamma_{\rm ren}$ as a function of $m$. 


\begin{thebibliography}{12345}
\bibitem{idet}
G.~V.~Dunne, J.~Hur, C.~Lee and H.~Min, Phys.\ Rev.\ Lett.\  {\bf 94}, 072001 (2005);
  Phys.\ Rev.\ D {\bf 71}, 085019 (2005).

\bibitem{radial1}
G. V. Dunne, J. Hur and C. Lee,
Phys. Rev. D {\bf 74}, 085025 (2006).


\bibitem{radial2}
G. V. Dunne, J. Hur, C. Lee and H. Min,
Phys. Rev. D {\bf 77}, 045004 (2008).


\bibitem{dunnekirsten}
  G.~V.~Dunne and K.~Kirsten,
  J.\ Phys.\ A. {\bf 39}, 11915 (2006).

  
\bibitem{gy}
  I.~M.~Gelfand and A.~M.~Yaglom,
  J.\ Math.\ Phys.\  {\bf 1}, 48 (1960);
K.~Kirsten and A.~J.~McKane,
  Annals Phys.\  {\bf 308}, 502 (2003).

  
\bibitem{hurmin}
  J.~Hur and H.~Min, Phys.\ Rev.\ D {\bf 77}, 125033 (2008).
  

\bibitem{thooft}
G.~'t Hooft,
  Phys.\ Rev.\ D {\bf 14}, 3432 (1976);
  [Errata {\it ibid}.\ D {\bf 18}, 2199 (1978)].


\bibitem{kwon}
  O.~K.~Kwon, C.~Lee and H.~Min,
  Phys.\ Rev.\ D {\bf 62}, 114022 (2000)

\bibitem{hurjin}  
 J.~Hur, C.~Lee and H.~Min, Phys.\ Rev.\ D {\bf 80}, 105024 (2009) [Arxiv:hep-th/0909.5515]
 

\end{thebibliography}
\end{document}